\documentclass[aps,prl,reprint,showpacs,showkeys,superscriptaddress,floatfix]{revtex4-1}
\usepackage{multirow}
\usepackage{graphicx}
\usepackage{bm}
\usepackage{dcolumn}
\usepackage[colorlinks=true,linkcolor=blue,urlcolor=blue,citecolor=blue]{hyperref}
\usepackage{natbib}
\usepackage{amsmath}
\usepackage{times}
\usepackage{xcolor}
\usepackage{soul}
\usepackage[normalem]{ulem}

\begin{document}

\preprint{2}

\title{Excess specific heat of the gapped sliding phonons in the incommensurate composite crystal Sr$_{14}$Cu$_{24}$O$_{41}$}

\author{Rabindranath Bag}\affiliation{Indian Institute of Science Education and Research, Pune 411008, Maharashtra, India}
\author{Soumitra Hazra}\affiliation{Indian Institute of Science Education and Research, Thiruvananthapuram 695551, Kerala, India}
\author{Rajeev Kini}\affiliation{Indian Institute of Science Education and Research, Thiruvananthapuram 695551, Kerala, India}
\author{Surjeet Singh}\email[email:]{surjeet.singh@iiserpune.ac.in}\affiliation{Indian Institute of Science Education and Research, Pune 411008, Maharashtra, India} 
\date{\today}
\begin{abstract}

We show that low temperature specific heat (C$_p$) of the incommensurate chain-ladder system Sr$_{14}$Cu$_{24}$O$_{41}$ is enriched by the presence of a rather large excess contribution of non-magnetic origin. Diluted Al doping at the Cu site or annealing the crystal in an O$_2$ atmosphere suppresses this feature considerably. Using the THz time-domain spectroscopy, we show that the occurrence of excess specific heat is associated with the presence of very low-energy ($\sim$ 1 meV) gapped phonon modes that originate due to the sliding motion of oppositely charged mutually incommensurate chain and ladder layers.   

\end{abstract}


\maketitle

The phononic ground state of crystalline solids is characterized by three gapless acoustic modes \cite{AshcroftandMermin}. This fundamental property is however vitiated in some composite crystals comprising two or more mutually incommensurate (IC) substructures. In these systems, new normal modes are predicted to emerge due to relative sliding motion of the IC substructures past each other \cite{Theodorou1978,Axe1982,Zeyher,Finger1983,Walker1985}. An extra longitudinal acoustic or acousticlike mode has indeed been observed experimentally in a number of IC systems, including Hg$_{3-x}$AsF$_6$ \cite{Heilmann1979, Hastings}, higher manganese silicides of the Nowotony chimney-ladder family \cite{ChenNaturecomm}, Rb-IV \cite{Loa}, n-alkane-urea composite crystals \cite{Toudic}, and in some of the BSCCO superconductors \cite{EtrillardEPL2001, EtrillardEPL2004}. A particularly interesting scenario arises when the IC substructures are oppositely charged rigid layers that slide past each other under the coulomb restoring force, which opens an energy gap rendering the sliding mode quasiacoustic or optical \cite{Theodorou1978}. These systems are interesting not only because they offer a tunable phonon gap which scales with the charge difference between the sliding layers, they also provide a platform for understanding gapped excitations in a wider class of IC or quasicrystals, including, disorderd solids and glasses \cite{Etrillard1996, Remenyi2015}.   
  
\begin{figure}[b]
	\centering
	\includegraphics[width = 0.48 \textwidth]{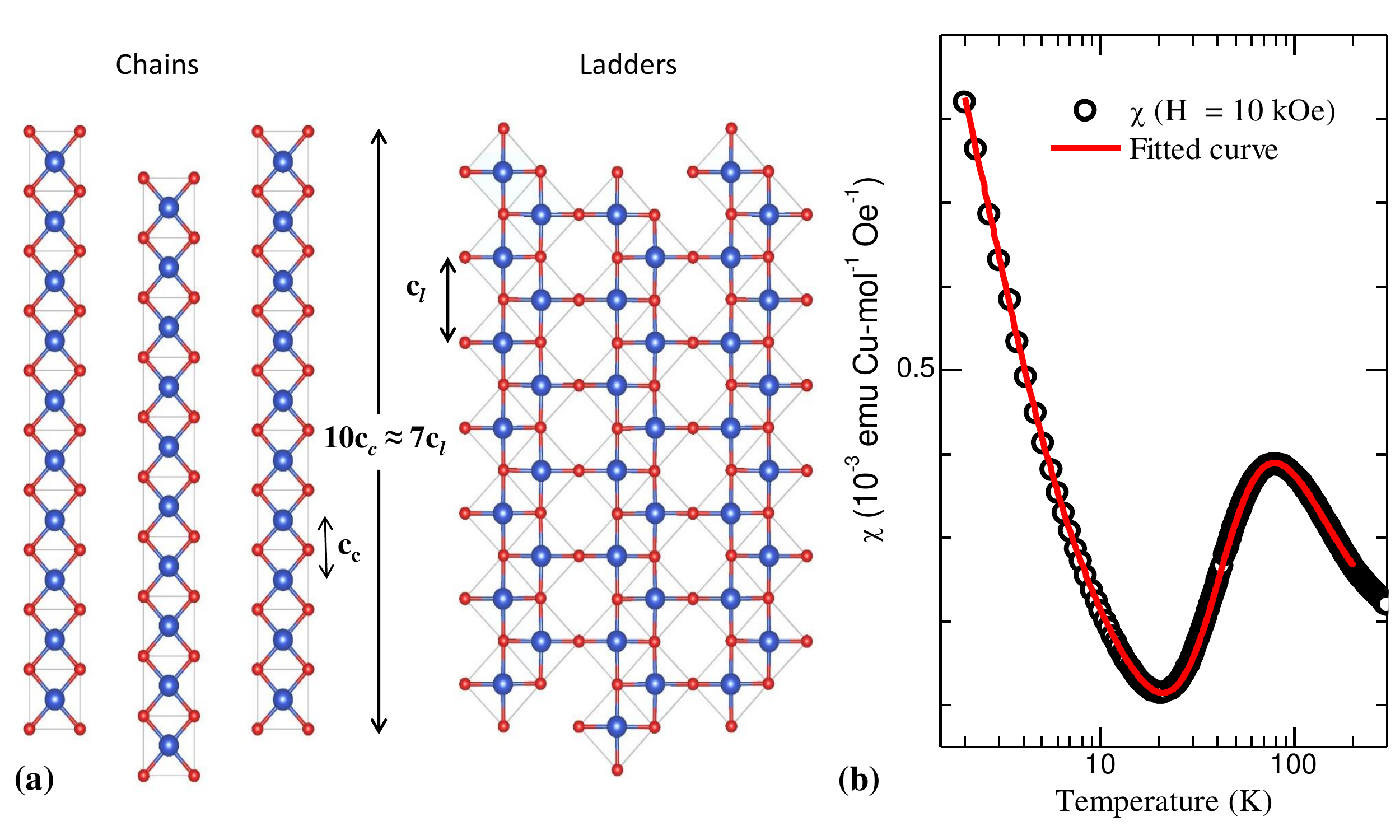}
	\caption{(a) The chain (CuO$_2$ ) and ladder (Sr$_2$Cu$_2$O$_3$) sublattices. Blue and red spheres represent Cu and O ions. The Sr ions are not shown for clarity. (b) Magnetic susceptibility $\chi$ of Sr$_{14}$Cu$_{24}$O$_{41}$ shown as a function of temperature. The measurements are done in external magnetic field of 10 kOe applied along the c-axis. The red curve is a best-fit to the data obtained using eq. \ref*{modchi} (see text for details).}
	\label{Pure_MT}
\end{figure}

In this regard, the chain-ladder compound Sr$_{14}$Cu$_{24}$O$_{41}$ is an ideal system for realizing the gapped sliding modes. It has a layered structure comprising an alternating stack of oppositely charged Sr$_2$Cu$_2$O$_3$ (positive) and CuO$_2$ (negative) layers. In the Sr$_2$Cu$_2$O$_3$ layer, the Cu ions form a network of weakly interacting 2-leg ladders running parallel to the c-axis; and in the CuO$_2$ layer they arrange as linear chains oriented parallel to the c-axis (Fig. \ref{Pure_MT}(a)). These layers are however mutually IC along the c-axis with $\alpha$ = c$_l$/c$_c$ $\approx$ $\sqrt{2}$, where c$_c$ ($\sim$ 2.75 \AA) and c$_l$ ($\sim$3.93 \AA) are the lattice parameters in the chain and ladder sublattices, respectively \cite{Hiroi}.   

While the magnetic ground state and spin excitations of Sr$_{14}$Cu$_{24}$O$_{41}$ have been extensively investigated in the past \cite{Dagotto1999}, the unconventional nature of its phononic ground state started gaining attention only recently. Thorsm$\o$lle et al. \cite{Thorsmolle} used Terahertz time-domain spectroscopy and Raman techniques to show the occurrence  of new infrared and Raman active modes in the very low energy range between 1 to 2 meV. They conjectured that these modes arise due to the gapped sliding motion of the chain and ladder layers. More recently, new optical phonon modes polarized along the incommensurate axis are reported in the same energy range using inelastic neutron scattering \cite{Chen}. However, the thermodynamic evidence of these modes and their unambiguous assignment with the relative sliding motion of the chain and ladder layers has remained an open question.  

Since these low-energy modes are appreciably gapped they should have a definitive thermodynamic signature in the low-temperature specific heat where the phononic specific heat is expected to exceed the Debye T$^3$ prediction. In this letter we show that the low temperatures specific heat (C$_p$) of Sr$_{14}$Cu$_{24}$O$_{41}$ is enriched by the presence of a rather large excess contribution of non-magnetic origin near T$^\ast$ = 10 K. Using a combination of THz-TDS and specific experiments on three different crystals, namely, as-grown, O$_2$-annealed and Al-doped, we establish that the excess specific heat in Sr$_{14}$Cu$_{24}$O$_{41}$ originates from the relative sliding motion of chain and ladder layers predicted previously. We also show that the gap-size associated with these modes is highly sensitive to the stoichiometry. 

Our experiments are performed on high-quality single crystals grown using the traveling-solvent floating-zone technique under a 3 bar oxygen pressure. The details of crystal growth and structural characterizations are given in Ref. \cite{BagJCG}. Magnetic and specific heat measurements were performed using a Physical Property Measurement System (PPMS), Quantum Design (QD), USA. For Terhertz Time-Domain Spectroscopy (THz-TDS) experiments, the THz signals were generated using ultrafast ($\sim$80 fs) photoexcitations of low-temperature grown GaAsBi epilayers; and were detected using a photoconductive antenna. Measurements were performed by mounting the sample on the cold-finger of an optical closed-cycle refrigerator. Transmittance were obtained by taking the ratio between magnitude of fast-Fourier transformed sample and the reference amplitudes. 

Temperature dependence of spin susceptibility ($\chi$) for H $\parallel$ c is shown in Fig. \ref{Pure_MT}(b). Since the ladder sublattice has a spin-singlet ground state with the singlet-triplet energy gap exceeding $\sim$35 meV \cite{Magish, Eccleston}, the low-temperature $\chi$ is practically due to the chain sublattice. Sr$_{14}$Cu$_{24}$O$_{41}$ is self-hole doped with six holes per formula unit. If all six holes are assumed to be present in the chain subsystem, the ground state configuration of the chains can be written as \cite{MasahiTakigawa1998}:  $\dotsb$ $\circ$ $\circ$ $[{\uparrow \circ \downarrow}]$ $\circ$ $\circ$ $[{\downarrow \circ \uparrow}]$ $\circ$ $\circ$ $\dotsb$,  where $\uparrow$ ($\downarrow$) represents a Cu$^{2+}$ spin 1/2, $\circ$ a Zhang-Rice singlet \cite{ZhangRicesinglet}, and $[{\uparrow \circ \downarrow}]$ denotes a spin 1/2 dimer. For this ideal configuration, $\chi$ should decrease  to zero exponentially as T $\rightarrow$ 0. However, in any real Sr$_{14}$Cu$_{24}$O$_{41}$ crystal, a small fraction of hole transfer from chains to ladders leads to a small number of unpaired spins in the chains that cause $\chi(T)$ to increase at low temperatures (T $<$ 20 K) in a Curie-like manner. Recently, it has been shown that at still lower temperatures (T $<$ 5 K),  these unpaired spins, separated by hundreds of Cu sites along the chain, anti-align to form dimers due to a small antiferromagnetic interaction mediated via the intervening sites \cite{Sahling2015}. Therefore, $\chi$ below T = 200 K can be described using a dimer-dimer model represented by eq. (\ref{modchi}): 

\begin{equation}
\chi = 
\chi_{0} + \frac{2N_{A}g^{2}\mu_{B}^{2}}{k_BT}\sum_{i}^{} \frac{N_{d_i}}{(3+e^{{-\frac{J_i}{k_{B}T}}})} + \frac{C}{T}, 
\label{modchi}
\end{equation}

where, $\chi_0$ ($\sim$10$^{-5}$ emu/mol) represents a small temperature independent van-Vleck contribution; the second term is the chain dimer susceptibility where i = 1 represents the dimers $[{\uparrow \circ \downarrow}]$, and i = 2 represents the long-distance dimers described above. N$_A$ denotes the Avogadro number, g is the Land\'e factor; $\mu_B$, the Bohr magneton and k$_B$ is the Boltzmann constant; N$_{d_i}$ denotes the number of dimers per f.u., and J$_i$ is the intradimer coupling. The last term in eq. (\ref{modchi}) represents the Curie contribution due to the spins that remain undimerized down to T = 2 K. 

As shown in Fig. \ref{Pure_MT}(b), a satisfactory fit to $\chi(T)$ is obtained using eq. \ref{modchi}. The best-fit value of J$_1$ and J$_2$ is about 132 and 2 K, respectively, in good agreement with previous reports (see \cite{Klingeler} for J$_1$ and \cite{Sahling2015} for J$_2$). The number of free spins N$_s$, calculated from C, is $\sim$0.006/f.u., and N$_{d1}$ and N$_{d2}$ are 1.89/f.u. and 0.04/f.u. respectively. As a passing remark, we should point out that fitting the data below T = 5 K without considering the low-temperature dimer term does not give a satisfactory fit at low-temperatures \cite{Bag2018}. 

 \begin{figure}[t]
	\centering
	\includegraphics[width = 0.5 \textwidth]{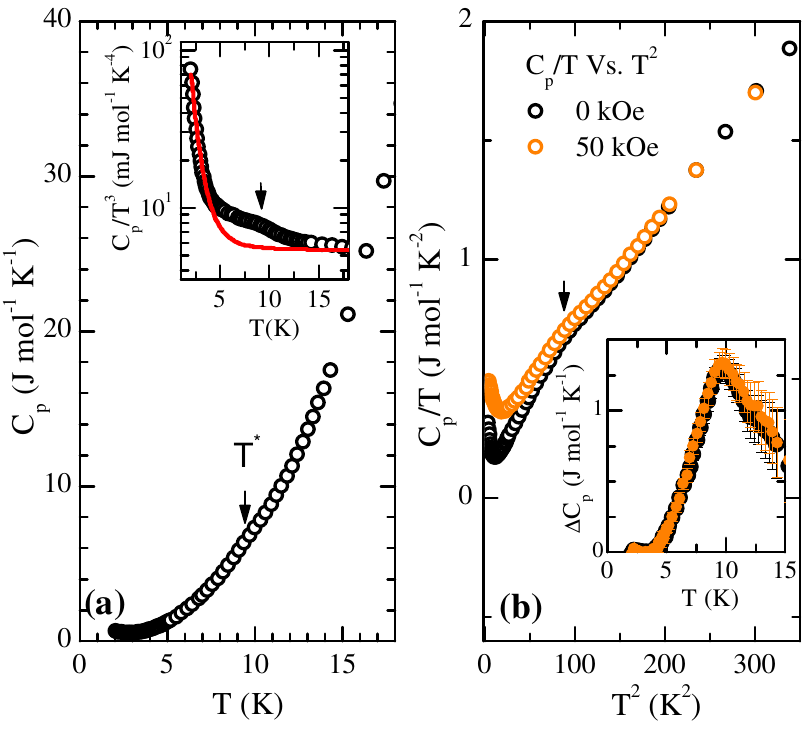} 
	\caption{(a) Specific heat (C$_p$) of the as-grown Sr$_{14}$Cu$_{24}$O$_{41}$ crystal plotted as function of temperature. (b) C$_P$/T vs T$^2$ in zero-field and under a field of 50 kOe. Inset in (a) show C$_p$/T$^3$ vs T. The solid red curve is a fit to the data in the range T = 2 to 4 K using eq. \ref{Spheat} (see text for details). Inset in (b) shows $\Delta$ C$_p$ plotted as a function of temperature for H = 0 and 50 kOe.} 
	\label{Pure_CT}
\end{figure}

The main result of our paper is shown in Fig. \ref{Pure_CT} where the specific heat (C$_p$) is shown as a function of temperature for the as-grown crystal. When plotted as C$_p$ Vs. T (panel a), the data show a fairly smooth variation except for the presence of a small blip near T$^\ast$ = 10 K. When re-plotted as C$_p$/T Vs. T$^2$ (panel b), this feature becomes more discernible. Additionally, now a low-temperature upturn, which marks the onset of long-distance dimerization is also visible. C$_p$/T is also shown in the presence of an applied magnetic field of 50 kOe. The field enhances C$_p$ at low-temperatures, but near T$^\ast$ the effect of magnetic field is relatively small. To determine the effect of field on C$_p$ near T$^\ast$, the low-temperature data away from T$^\ast$, in the temperature range from T = 2 to 4 K, is fitted using the spin-dimer model (eq. (\ref{Spheat})): 

\begin{equation}
\label{Spheat}
C_{p}(T) = RN_{d_2}(J_2/k_BT)^{\frac{3}{2}} e^{- J_2/k_BT} + \beta T^3 + AT^{-2},
\end{equation}

here, the first term represents the contribution of long-distance dimers ($R$ here is the gas constant), $\beta$T$^3$ is the phononic contribution, and the term A/T$^2$ represents the Schottky-tail due to a small number of unpaired spins, i.e., the spins that remain undimerized down to 2 K. The fitted curve (shown in the inset of Fig. \ref{Pure_CT}(a) where C$_p$ is plotted as C$_p$/T$^3$ vs. T) is extrapolated to higher temperatures. The excess specific heat ($\Delta$C$_p$), estimated by taking the difference between the experimental data and the fitted curve, is shown in the inset of Fig. \ref{Pure_CT}(b) for H = 0 and 50 kOe. $\Delta$C$_p$ exhibits a well-defined peak, and its non-magnetic origin can be inferred from the absence of any anomalous feature in $\chi(T)$ near T$^\ast$, and the insensitivity of $\Delta$C$_p$ to an applied magnetic field. The peak height exceeds 1 J mol$^{-1}$ K$^{-1}$ at T$^\ast$, which is rather large given its non-magnetic origin. A comparable excess C$_p$ is previously reported in the pyrochlore Bi$_2$Ti$_2$O$_7$ where the quenched configurational disorder among the Bi lone pairs results in departure from the Debye T$^3$ law \cite{Melot}. 

\begin{figure}[!]
	\centering
	\includegraphics[width = 0.41 \textwidth]{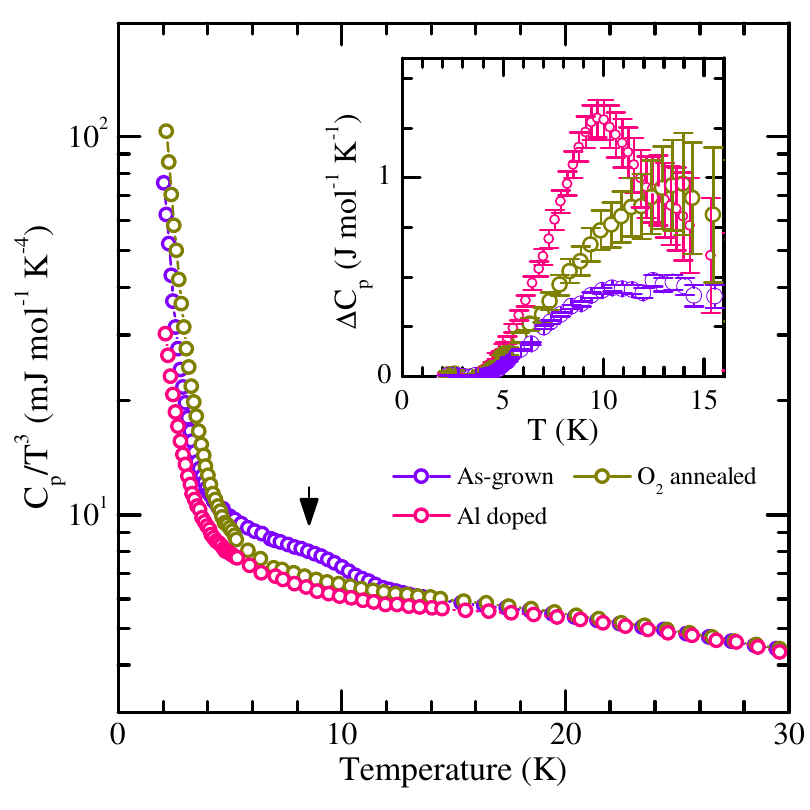} 
	\caption{Specific heat (C$_p$) of an as-grown, O$_2$ annealed and Al-doped crystals. Inset shows $\Delta$C$_p$ for the same three crystals plotted as a function of temperature.}
	\label{Doped_CT}
\end{figure}

In Fig. \ref{Doped_CT}, C$_p$ is shown for the O$_2$ annealed and 0.25\% Al doped crystals. The annealing treatment was carried out in a tubular furnace under flowing O$_2$ at T = 850 $^\circ$C for 36 hrs, and the doping concentration of Al mentioned here refers to the actual value obtained using the inductively coupled plasma technique. Below T = 4 K, where the long-distance dimerization of the unpaired spins is expected to set-in, C$_p$ is enhanced due to O$_2$ annealing but suppressed for Al doping. However, in both cases, C$_p$ exhibits a strong suppression in the region around T$^\ast$, which is what we are primarily interested in. Using the procedure described above, $\Delta$C$_p$ for O$_2$ annealed and Al doped sample is extracted and compared with that of the as-grown crystal as shown in the inset of Fig. \ref{Doped_CT}. Upon annealing or lightly doping, the anomaly in $\Delta C_p$ decreases in magnitude and shifts towards higher temperatures. 

\begin{figure}[!]
	\centering
	\includegraphics[width = 0.5 \textwidth]{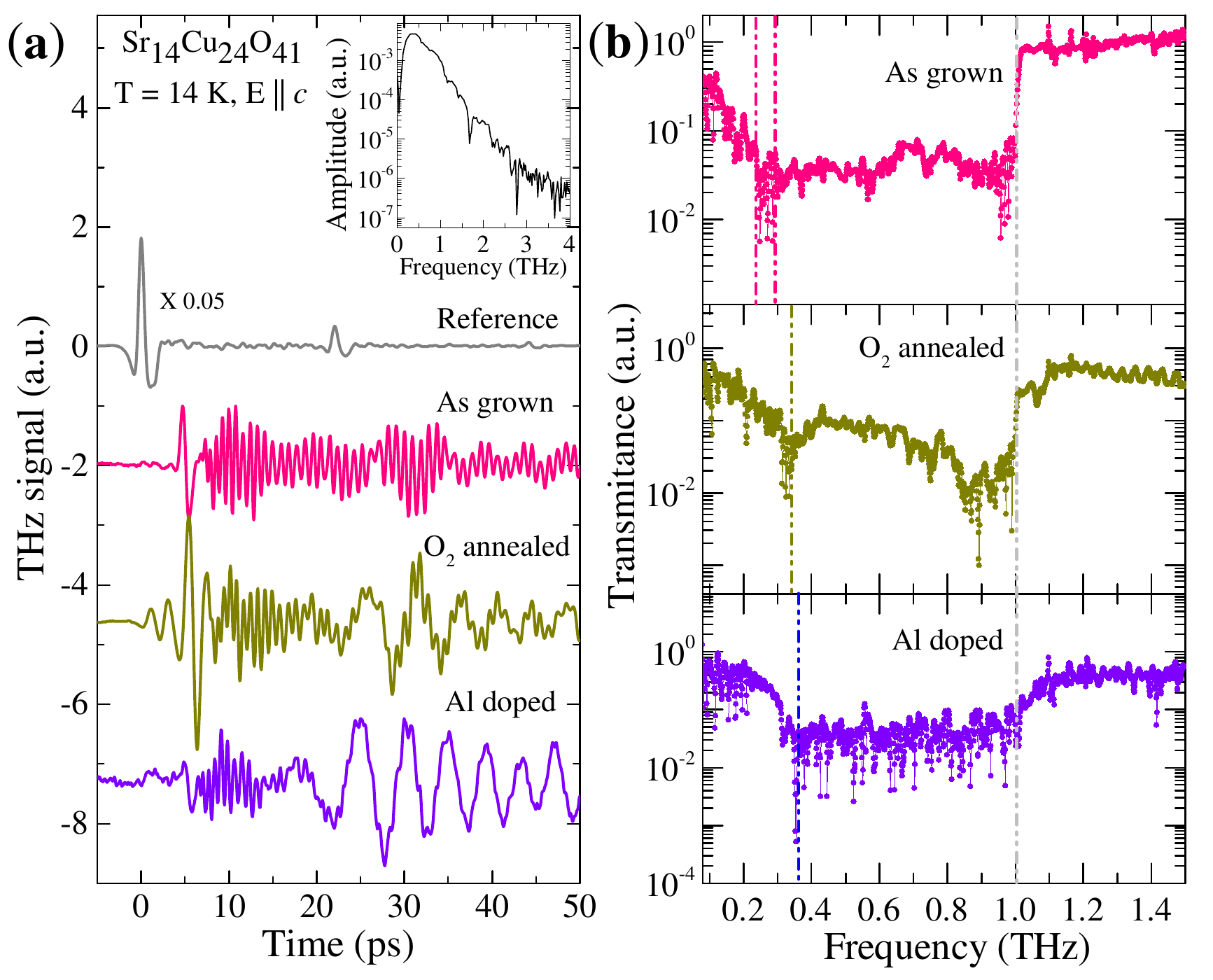}
	\caption{(Left panel) Transmitted THz electric field (E) at T = 14 K and E $\parallel$ c; (Right panel) THz transmittance for E $\parallel$ c at T = 14 K for the as-grown, oxygen  annealed (O$_2$ ann) and Al-doped Sr$_{14}$Cu$_{24}$O$_{41}$ crystals. Inset in the left panel shows fast-Fourier transform amplitude spectrum of the THz electric field for the reference.}
	\label{THz}
\end{figure}

In Fig. \ref{THz}(a), the THz signal transmitted through the sample is shown for the as-grown, oxygen annealed and Al-doped crystals at T = 14K and with electric field (\textbf{E}) polarized along the c-axis (E $\parallel$ c). The reference signal without the sample is also shown. The THz electric field of the as-grown crystal can be broadly characterized by the presence of two types of oscillations having periods $\sim$1 and $\sim$4 ps. For E $\parallel$ a (not shown), the THz signal showed no oscillations in line with the data reported earlier by Thorsm$\o$lle et al. \cite{Thorsmolle}. As shown in fig. \ref{THz}, for E $\parallel$ c, the annealed and doped crystals show a significantly different behavior compared to the as-grown crystal. To assess this difference quantitatively, THz transmittance for all three samples is shown in Fig. \ref{THz}(b). The transmittance drops by more than an order of magnitude within a sharply defined window between $\sim$0.3 THz to $\sim$1THz (1 THz = 1 ps = 33.3 cm$^{-1}$ = 4.1 meV) in agreement with Ref. \cite{Thorsmolle}. The lower and upper cut-offs near 0.3 and 1 THz correspond to the slow and fast oscillations, respectively.      

As mentioned earlier, Sr$_{14}$Cu$_{24}$O$_{41}$ has an IC structure with oppositely charged layers whose relative sliding motion gives rise to new gapped phonon modes. IR modes associated with the sliding motion were previously reported by Thorsm$\o$lle et al. at the lower cut-off (0.3 THz) of the THz transmittance window \cite{Thorsmolle}. In the inelastic neutron scattering investigations also low-lying optical phonon modes were revealed in the same energy range \cite{Chen}. 

In the THz transmittance data of our as-grown crystal, we see an absorption band near the lower cut-off in the frequency range 0.25 to 0.30 THz. We attribute this to the sliding motion based on the previous study \cite{Thorsmolle}. Interestingly, upon annealing the crystal, this band became narrower and its center shifted to a higher frequency at 0.35 THz. To understand this dramatic change upon mild annealing of the crystal, we recall that the degree of incommensurability in the IC systems depends sensitively on the sample's stoichiometry \cite{Smaalenreview}. In the present case, the stoichiometry can be written down in terms of $\alpha$ as: [CuO$_2$]$_{\alpha}$[Sr$_2$Cu$_2$O$_3$], where $\alpha$ = 10/7 for an ideal formulation. However, as shown by Hiroi et al., the as-prepared Sr$_{14}$Cu$_{24}$O$_{41}$ samples exhibit a broad distribution of $\alpha$ values around the commensurate point $\alpha$ = 10/7, which implies that the as-prepared samples are expected to exhibit a high and varying degree of incommensurability. Interestingly, upon oxidizing or reducing the sample, which either adds or remove oxygen from the chains or ladders thereby changing $\alpha$, the distribution width can be narrowed down considerably \cite{Hiroi}. We believe that oxygen annealing has a similar effect in our crystal; i.e., it minimizes or reduces the degree of misfit between the chains and ladders, which results in suppression of the excess specific heat associated with sliding motion that results from this misfit.    

As a consistency test we checked if small changes in $\alpha$ can indeed stiffen the mode frequency appreciably or not. For this we estimated the change $\delta \omega$ in $\omega$ for a small value of $\Delta\alpha$. The frequency $\omega$ of the sliding mode for a system with incommensuarte, oppositely charged layers is given by \cite{THEODOROU1980}: $\omega^2$ = $\omega_0^2$(1 + $\mu \alpha$), where $\mu$ is the ratio of the unit cell masses of the two sublattices; $\omega_0$ is analogous to the plasma frequency in metals and is given by $\omega_0^2 = n_c q_c^2/\epsilon \epsilon_0 m_c$, where n$_c$ is atomic mass number density and m$_c$, the unit cell mass (subscript c is for the chain sublattice); $\epsilon$ is the permittivity which can be taken as 15 \cite{Thorsmolle}. Using this expression, and by taking a reasonably small and experimentally accessible value of $\delta \alpha$ = 0.01, and the average frequency $<\omega>$ = 0.28 THz, we found $\Delta \omega$ = $\sim$0.03 THz, which is in fairly good agreement with the shift of the absorption band in the THz transmittance upon annealing.       

The effect of Al doping is more complex as it not only affects the oxygen stoichiometry as discussed next but also acts as a pinning center owing to its different ionic radius and charge state compared to the host ion. Since Al concentration is very small, to understand its effect on the sliding motion, we first consider only the effect of Al doping on the oxygen stoichiometry. Since Al$^{3+}$ replaces Cu$^{2+}$ in the lattice, the charge neutrality requires an equivalent quantity of oxygen absorption by the sample unless some Cu$^{2+}$ reduces to Cu$^{1+}$ which is unlikely due to an oxygen rich growth atmosphere employed during the growth experiment. Therefore, in the first approximation, the effect of Al doping is similar to oxygen intake due to annealing discussed in the preceding paragraph. However, the smaller ionic radius and higher charge state of Al$^{3+}$ is also expected to contribute by pinning of the sliding DW in a manner analogous to impurity pinning in charge and spin DW systems. The upper cut-off of the transmittance near $\sim$4 meV energy (1 THz), which we did not discus in this paper, also show much higher sensitivity to Al doping relative to O$_2$ annealing for which the cut-off remains sharply defined and unshifted. Since the charge/spin ordering in the chains remains unaltered upon small Al-doping \cite{Bag2018}, these changes possibly reflect the corresponding modulations of the charge density wave (CDW) pattern in the ladders upon Al doping.        

Let us now discuss the origin of excess C$_p$ at T$^\ast$ and its subsequent suppression upon lightly doping or annealing the sample. The temperature variation of $\chi$ shows no anomalous feature around T$^\ast$ and it is well-fitted using the dimer-dimer model, which suggests a non-magnetic or likely phononic origin of the excess C$_p$. This observation is also consistent with the fact that $\Delta C_p$ is field independent. The conventional low-dimensional CDW systems \cite{george2000density}, for example,  K$_{0.3}$MoO$_3$ \cite{Odin2001}, (TaSe$_4$)I \cite{BILJAKOVIC1986PRL}, also show excess specific heat at low temperatures but this arises due to the collective phase (phason) and amplitude (amplitudon) modes of the IC modulation below the CDW transition \cite{Remenyi2015}. Though Sr$_{14}$Cu$_{24}$O$_{41}$ also undergoes a CDW ordering in the ladders below T $\approx$ 200 K, the CDW in this case is not accompanied  by a lattice distortion, and is believed to arise from the many-body electronic effects \cite{Abbamonte}. Therefore, a similar scenario as operative in conventional CDW systems seems inconceivable in the present case. Our experimental results suggest that the excess specific heat in Sr$_{14}$Cu$_{24}$O$_{41}$ arises from the relative sliding motion, which is manifested in the form of THz modes near the lower cut-off of the transmittance window. The temperature T$^\ast$ = 10 K of peak in $\Delta$C$_p$, and an almost exponential rate of decrease at lower temperatures are consistent with a gapped scenario with a gap size close to 1 meV. C$_p$ and THz of the annealed and Al doped crystals also lend support to this interpretation. In the annealed sample, the peak in $\Delta$C$_p$ shifts to higher temperatures from $\sim$10.5 K to $\sim$13.5 K, this is in line with the observed stiffening of the lower cut-off from $\sim$0.28 THz (average $\omega$ for the as-grown crystal) to 0.35 THz upon oxygenation. In the case of Al-doped sample, the stiffening of the lower cut-off is even more pronounced, which is not unexpected since Al impurities should also act as pinning centers impeding the sliding motion.        

To conclude, we showed the presence of a large excess specific heat in the IC chain-ladder compound Sr$_{14}$Cu$_{24}$O$_{41}$ around T$^\ast$ = 10 K. However, no magnetic anomaly could be detected in the $\chi(T)$ data in this temperature range, which suggests that the excess C$_p$ is of non-magnetic origin. The insensitivity of $\Delta$C$_p$ to the applied magnetic field further corroborates this conclusion. By using THz-TDS in the range from $\sim$0.2 to $\sim$1.4 THz, we argued that the excess C$_p$ originates from the sliding phonon modes. Experiments on O$_2$ annealed and Al doped crystals revealed significant suppression of the excess C$_p$, and concomitant changes in the THz response associated with the sliding modes. A very high sensitive of THz response to minor structural changes suggest that  Sr$_{14}$Cu$_{24}$O$_{41}$ has a highly unconventional phononic ground state which invites further investigations.

\begin{acknowledgements}

\end{acknowledgements}

\bibliographystyle{apsrev4-1}
\bibliography{Boson}

\end{document}